# Improving ecological niche models by data mining large environmental datasets for surrogate models

David R.B. Stockwell

Address for correspondence:

*San Diego Supercomputer Center,*
*University of California, San Diego*
*9500 Gilman Drive, La Jolla, CA 92093-0505*
*ph: +1 858 8220942, fax: +1 858 8223631, davids@sdsc.edu*

## Abstract

WhyWhere is a new ecological niche modeling (ENM) algorithm for mapping and explaining the distribution of species. The algorithm uses image processing methods to efficiently sift through large amounts of data to find the few variables that best predict species occurrence. The purpose of this paper is to describe and justify the main parameterizations and to show preliminary success at rapidly providing accurate, scalable, and simple ENMs. Preliminary results for 6 species of plants and animals in different regions indicate a significant ($p<0.01$) 14% increase in accuracy over the GARP algorithm using models with few, typically two, variables. The increase is attributed to access to additional data, particularly monthly vs. annual climate averages. WhyWhere is also 6 times faster than GARP on large data sets. A data mining based approach with transparent access to remote data archives is a new paradigm for ENM, particularly suited to finding correlates in large databases of fine resolution surfaces. Software for WhyWhere is freely available, both as a service and in a desktop downloadable form from the web site http://biodi.sdsc.edu/ww_home.html.

## Introduction

Since the inception of ecological niche modeling, finding better methods of answering the question "Where is it and why?" has been a fundamental objective of modelers (Stockwell 1993). Choosing among the many forms of predictive models of habitat distribution in ecology has not been based on statistical performance in a single trial, but has included the objectives of the study and generality (Guisan and Zimmermann 2000). For example, it has been shown that different forms of models may be more accurate at different sample sizes (Stockwell and Peterson 2002) although ecological niche models (ENMs) developed by in GARP (Genetic Algorithm for Rule-set Prediction) are accurate over a range of sample sizes (Stockwell and Peters 1999). However, the 'why' question is multi-faceted and not easily inferred from 'black-box' complex models (Stockwell et al., 1990). A stepwise removal procedure indicates critical variables in GARP rule-sets (Peterson and Cohoon 1999), but this approach is time-consuming and provides little additional information. Two segmented networks have been used to overcome the 'black box' problem and explain how confident a neural net is in its conclusions (Werner and Obach 2001). Generalized linear models (GLMs) and Generalized Additive Models (GAMs) explain by nature of their functional forms representing the uni-modal expectations of ecological niche theory (Austin 2002).

It is not clear that agreement with ecological theory – e.g. the principle of central tendency in niche theory - is sufficient to confidently answer 'why'. For example, the use of linear instead of non-linear models leads to sub-optimal species distribution models (Austin and Meyers, 1990, Austin 2002), and climatic envelopes defined by the Mahalanobis distance are more accurate that rectilinear envelopes (Farber and Kadmon 2003). However, highly non-Gaussian distributions are problematic for all parametric statistical methods (James and McCulloch, 1990). The distribution of environmental values are almost always highly skewed, eg. temperatures are generally moderate throughout most of the landscape, with small areas of extreme cold on mountain summits. In addition, extreme non-linearites not representable by uni-modal functions are common, particularly in remote sensing data. For example, values of the percent of vegetation cover in the continuous fields data (Hanson et al., 2003) range from 20 to 80, but zero cover has value 255. These data sets would not perform well in models of central tendency due to inversion of the natural placement for the zero value. The ideal ENM method will (1) be capable of modelling a wide range of responses, (2) allow critical examination of assumptions, and (3) be a simple approach that will not fit inappropriate functions, but (4) will handle extremely non-linear data, and (5) will efficiently turn an increasing flood of data from satellites, geographic information systems and climate model outputs into simple, scalable ENMs.

In a new approach to this problem, we describe the WhyWhere algorithm, which integrates a dynamic categorization procedure with a form of data mining, sifting through large amounts of data with an efficient image processing routine to discover accurate 'surrogate' models. A surrogate model in ENM is a one dimensional (1D) model with a discrete categorization of the landscape, such as an aerial photographic or satellite maps of land classes based on vegetation. Surrogate, refers to the way landcover maps stand in place of species habitat (Stomes and Estes, 1993). Developers calibrate surrogates for prediction with the frequency with which survey points fall within the classes. The surrogate approach is simple, intuitive and has figured highly in recent work on the prediction of species distributions and understanding patterns of biodiversity (e.g. Scott et al., 1996). Difficulties in obtaining adequate survey data, high-resolution vegetation maps, and concerns with the statistical validity have tended to limit their application (Stockwell and Peterson, 2003). However, in a comparative study with a large number of species, simple 'surrogate' maps generally equaled or exceeded accuracy of other multivariate methods (Stockwell and Peterson, 2002). WhyWhere exploits a natural analogy between a landcover map and the colors in an image, developing a surrogate model by converting between 3D 'raw' formats, where the intensity of each color is stored for each pixel, and a 1D 'palette' format (e.g. GIF), a size-limited list of colors. The color reduction algorithm called Heckbert's median cut (Heckbert, 1982) converts a continuous color image to a reduced color-categorized image. This algorithm is widely regarded as providing a natural appearance to the human eye with high data compression. This algorithm divides the pixels in the image into equally sized categories. More categories are assigned to large gradually changing color areas (e.g. flesh tones, green vegetation) than to small areas of color (e.g. a small red ball, a mountain top). The median cut approach and discrete binning of colors in Heckbert's algorithm inherently handles extremely skewed color distributions. Software for WhyWhere is freely available, both as a service and in a desktop downloadable form from the web site http://biodi.sdsc.edu/ww_home.html.

## *Method*

Model development follows the usual stages of species distribution modeling (Stockwell and Peters, 1999). There are a number of enhancements for ease of use however, particularly transparent access to environmental data from a large remote data archive.

**Preparation of environmental data.** Point data are entered as longitude latitude (x y) pairs. A set of environmental layers are prepared by cropping and scaling global extent data sets to the same resolution and extent as the point data. The global data sets may be stored in a remote archive and accessed transparently to the user (providing they are on the internet). Alternatively, global data may be locally, or copied into the working directory. Regions of the image not of interest (such as water in a terrestrial analysis) are masked out. Point data are then filtered to produce one unique point per grid cell, and the geographic extent is determined from the range of the point data.

**Model development and testing;** At each iteration the algorithm generates new surrogate models from the combination of each variable with the most accurate variable in the previous channel(s), stopping when a new channel does not significantly exceed the accuracy of the previous channels. The most accurate model is the image at the penultimate iteration, e.g.

```
1. Find the most accurate environmental variable.
2. Put this variable in the red channel of an image and find the
   most accurate combination with each of the environmental files in
   the green channel.
3. If necessary, test this image with each of the environmental
   layers in the blue channel.
```

Calibration of surrogate: In this case developing the model reduces to calibrating each category (color) for the conditional probability of a species. A simple heuristic on the number of data points auto-selects the number of colors *c* if required:

$$if\ n < 16\ then\ c = 16\ else\ c = 255$$

For a color *i*, where *P* is the proportion of each occurrence points of a color *i* and $B_i$ is the proportion of absences (or background) of a specific color then the expected probability *Pr* is given by the rule:

$$if\ (P_i+B_i) > 0,\ then\ Pr_i = \frac{P_i}{P_i+B_i}\ else\ Pr_i = 0$$

The probabilities in array *Pr* can predict the distribution using a cut *x* in a characteristic function $X{:}I{\rightarrow}\{0,1\}$:

***X(i): if Pr_i > x then 1 else 0***

Objective functions:

1. Maxent

2. The accuracy *A* is the sum over each *n* colors color divided by two (as $P_i$ and $B_i$ sum to one.)

$$A = \sum_{i=0}^{n} \frac{X(i)P_i + (1-X(i))B_i}{2}$$

The difference in accuracy between models provides a measure of significance, and can be calculated using the *Z* statistic, assuming the binomial distribution, generated from the independent sampling of each point in the populations of presences and absences. While this assumption may be violated by spatial autocorrelation, the algorithm has attempted to reduce this by eliminating duplicate occurrences and random sampling of data points. The *Z* statistic for the difference in accuracy *A* of models where *n* is the number of points is:

$$Z = \frac{|A_1 - A_2|}{\sqrt{\frac{A_1 A_2}{n}}}$$

**Prediction to new areas**: For extrapolative predictions where a model developed in a specific environmental data set is applied in different data set, either a different location (invasive species) or with different base layers (migration, climate change) the list of colors, is used to categorize a new multi-channel image into the original set of colors. This operation can be performed in *netpbm* tools with any of *pnmcolormap*, *pnmremap* or *pnmquant*.

**The explanation stage;** allows study of the response surface and conversion to a variety of formats for visualization and mapping. The species distribution is mapped by changing the palette colors in an image according to the probability: e.g. red is highly probable, green is low and blue is zero. The response surface of the points of presence and absence can be shown in a histogram which is two dimensional in the case of one variable, and 3 dimensional in the case of models with more than one variable.

The data sets consisted of publicly available, raster formatted environmental datasets in geographic (lat. long.) projection, with global coverage drawn from the EPA Global Ecosystems Dataset v2.0 (NOAA-EPA, 1992). Additional satellite data such as the NDVI continuous fields data with a 1 km resolution, were obtained from the Global Land Cover Facility (Hanson et al., 2003). Table 1 contains a selection to demonstrate the current range of data sets and their resolutions. Occurrence points for six species on 6 species of birds and plants (Cerulean Warbler, Swainson's Flycatcher, Eared Trogon,

Rubber Vine, and two user supplied mammal data sets) in six regions of the world (North America, South America, Mexico, Australia, Veracruz, and Brazil) respectively, test and evaluate parameter settings for the algorithm. External accuracy is the only reliable measure of accuracy, as models with high internal accuracy may be 'overfitting' the data, being too specialized and consequently performing poorly on new data (Verbyla and Litvaitis, 1989). In this study, 80% of data were used to develop the model and 20% were held back for evaluating external accuracy.

### *Results*

Figure 1 shows the internal and external accuracy with Cerulean Warbler with different numbers of data, showing the typical convergence of internal and external accuracy as the number of data points increases, as found in all modeling algorithms (Stockwell and Peterson, 2002). Fig 2 shows the response surface for Cerulean Warbler from WhyWhere: the number of points at each color, covering the range of average annual precipitation (lwcpr00). The response surface illustrates critical features of an ecological niche model: the distribution of the presence points is restricted to a range; the background points cover the whole range, and presence of an outlier and the non-linearity of the response.

**Channels**: Is the restriction to three variables (red, green and blue channels) adequate to optimize accuracy? Fig 3 shows the internal and external accuracy of models at each iteration of the algorithm on three test species. Both the internal and external accuracy generally maximizes in two iterations. Thus the simpler strategy of selecting two variables using internal accuracy may be adequate.

**Categories**: Fig 4 shows the accuracy on a range of different size Cerulean Warbler data sets for 255, 128 and 64 categories. The 255 categories produced higher external accuracies for all data sets except for the smallest. Thus 255 categories is adequate, but optimal choice of numbers of categories is important on small numbers of data.

**Accuracy**: We compared WhyWhere with GARP under typical usage conditions. GARP used annual average climate, elevation, and vegetation datasets described in Stockwell and Peterson (2003), and WhyWhere had access to a suite of data sets from Table 1. The sampling protocol was identical, with the data selected to an initial prior probability of 0.5, and testing carried out on random samples with replacement. The accuracy of GARP was $0.77 \pm 0.03$, and WhyWhere has $0.88 \pm 0.03$, a highly significant ($p<0.01$) 14% increase.

**Speed**: At small map sizes (coarse resolution), it is slowed by constant aspects of the algorithm but at finer resolutions and larger data sets WhyWhere was 583% faster than GARP higher resolution of distributions feasible (Figure 5).

**Resolution**: The Cerulean Warbler was predicted at spatial resolutions from 1 deg. to 0.05 deg. (Table 3). Three variables were chosen repeatedly for the first channel:

mgv0009 (Average September Generalized Global Vegetation Index), lcprc04 (Leemans and Cramer April Precipitation) or lwcpr04 (Legates & Willmott April Corrected Precipitation) and the second channel varied. The accuracy increased slightly with increasing resolution (0.87 to 0.89) but was not significant.

## *Discussion*

The WhyWhere system shows many desirable characteristics. Firstly, there is a significant increase in accuracy and speed over GARP. Secondly, the models are typically composed of few (typically two) variables. Thirdly, the approach has desirable features for a modeling method in a generic, analytical information infrastructure. The capacity to crop and scale environmental data from large remote archive removes the need to develop region-specific base layers. If required the user can use their own data sets, or contribute data to the archive. The expansion of potential data sets will increase the of applications beyond species distributions to other types of occurrences with other potential environmental correlates e.g.: geomorphology such as landslides, or social such as crime.

Fewer variables gave greater accuracy. If we observe Occams' razor, we should prefer the simpler model, as fewer variables implies better explanations than complex models with many variables, such as those produced by GARP or neural nets. In addition, the simple model of the response of the species shown in Fig 2 illustrates the classical univariate response also seen in GLMs and GAMs, and the potential to represent more extreme non-linearities such as outliers, out of order values, and skewed and multi-modal responses. Improvements in both prediction and explanation have implications for one of the most vexing questions in ecological modeling: the difficulty of simultaneously developing models that both predict and explain (Loehle, 1983; Stockwell, 1993). Thus, in the future, with the robust non-linearity of the analysitical approach, WhyWhere could become a general-purpose predictive and explanatory system to enable new research and development directions.

A larger evaluation study found surrogate, logistic regression and GARP methods gave similar accuracy standardized protocol (Stockwell and Peterson, 2002), suggesting the increase in accuracy is due to access to a greater range of data rather than inherent in the model itself. Examination of the variables selected by the system for the 6 species (Table 2) shows *treecover* is selected most frequently (3 times), while *mgvc188,* the first eigenvector of vegetation, is selected next most frequently (2 times) and the rest were monthly climate variables. As the data sets used in GARP were annual averages of climate and vegetation, improvements in the accuracy of WhyWhere could be attributed to the monthly climate data sets (i.e. greater temporal resolution). One area for improvement is in the heuristic for binning environmental data. The number of categories affects accuracy of surrogate models (Stockwell and Peterson, 2002). Optimizing this choice is important, as one of the main sources of generalization in the method is the choice of categories, and controlling overfitting and subsequent accuracy on independent test data.

## *Acknowledgements*

Point data for this study were kindly made available Townsend Peterson, Leo Joseph, Karen Stocks, and Denis Filer. Versions of the software for distribution were developed by Haowei Liu. Joseph Kirkebride provided suggestions on a later draft. This work was partially funded by the National Science Foundation primarily through the National Science Foundation grants SEEK: Science Environment for Ecological Knowledge (DBI0225674) and ITR: Collaborative Research: Building the Tree of Life – A National Resource for Phyloinformatics and Computational Phylogenetics (EF0331648).

## *References*

Austin, M., A. Nicholls, and Margules. C., 1990. Measurement of the realized qualitative niche - environmental niches of 5 Eucalypt species. *Ecological Monographs* 60: 161-177.

Austin, M.P., 2002. Spatial prediction of species distribution: an interface between ecological theory and statistical modeling. Ecological Modelling, 157, 101-118.

Davey, S.M., Stockwell, D.R.B., 1991. Incorporating habitat into an Artificial Intelligence framework: concepts, theory and practicalities. A.I. Applications in Natural Resource Management **5,** 59-104.

Farber, O., and Kadmon, R., 2003. Assessment of alternative approaches for bioclimatic modeling with special emphasis on the Mahalanobis distance. 160, 115-130.

Guisan, A., and Zimmermann, N.E., 2000. Predictive habitat distribution models in ecology. Ecological Modelling, 135, 147-186.

Heckbert, P., 1982. Color Image Quantization for Frame Buffer Display, SIGGRAPH '82 Proceedings, page 297.

James, F.C., McCulloch, C.E., 1990. Multivariate analysis in ecology and systematics: panacea or Pandora's box? Annual Review of Ecology and Systematics 21, 129-166.

Jarvis, A.M., Robertson, A., 1999. Predicting population sizes and priority conservation areas for 10 endemic Namibian bird species. Biological Conservation 88, 121-131.

Hansen, M.; DeFries, R.; Townshend, J.R.; Carroll, M.; Dimiceli, C.; Sohlberg, R., 2003. 500m MODIS Vegetation Continuous Fields. College Park, Maryland: The Global Land Cover Facility.

Loehle, C., 1983. Evaluation of theories and calculation tools in ecology. Ecological Modelling, 19, 239-247.

NOAA-EPA Global Ecosystems Database Project, 1992. Global Ecosystems Database Version 1.0. User's Guide, Documentation, Reprints, and Digital Data on CD-ROM. USDOC/NOAA National Geophysical Data Center, Boulder, CO.

Peterson, A.T. and Cohoon, K.P., 1999. Sensitivity of distributional prediction algorithms to geographic data completeness. Ecological Modelling, 117:159-164.

Scott, J.M., Tear, T.H., Davis, F.W., 1996. Gap Analysis: A landscape approach to biodiversity planning. American Society for Photogrammetry and Remote Sensing, Bethesda, Maryland. ISBN-1-57083-03603

Stockwell, D.R.B., Davey, S.M., Davis, J.R., Noble, I.R., 1990. Using induction of decision trees to predict Greater Glider density. A.I. Applications in Natural Resource Management 4, 33-43.

Stockwell, D.R.B., Noble, I.R., 1992. Induction of sets of rules from animal distribution data: a robust and informative method of data analysis. Mathematics and Computers in Simulation 33, 385-390.

Stockwell, D.R.B., 1993. Machine learning and the problem of predictions and explanation in ecology. PhD Thesis, Australian National University.

Stockwell, D.R.B., Peters, D., 1999. The GARP Modeling System: problems and solutions to automated spatial prediction. International Journal of Geographical Information Science 13, 143-158.

Stockwell, D.R.B., 1999. Genetic Algorithms II. In A.H. Fielding (Ed.), Machine Learning Methods for Ecological Applications. Kluwer Academic Publishers, Boston, pp 123-144

Stockwell, D.R.B., Peterson, A.T., 2002. Effects of sample size on accuracy of species distribution models. Ecological Modelling 148, 1-13.

Stockwell, D.R.B., Peterson, A.T., 2003. Comparison of resolution of methods for mapping biodiversity patterns from point-occurrence data. Ecological Indicators 3, 213-221.

Stomes, D.M., Estes, J.E., 1993. A remote sensing research agenda for mapping and monitoring biodiversity. International J. of Remote Sensing 1839-1860

Verbyla, D.L., Litvaitis, J.A., 1989. Resampling methods for evaluating class accuracy of wildlife habitat models. Environmental Management, 13 783-787.

Werner, H., and Obach, M., 2001. New neural network types estimating the accuracy ofr response for ecological modelling. Ecological Modelling, 146, 289-298.

# Tables

Table 1.  Examples of Terrestrial datasets in WhyWhere

| Name | Description | Resolution (degrees) | No of Vars |
|---|---|---|---|
| treecover | Continuous field data - treecover | 0.01 | 1 |
| wrzsoil | Webb et al Soil Particle Size Properties Zobler Soil Types | 1.0 | 1 |
| wrtext | Webb et al Texture-Based Potential Storage of Water (mm) | 1.0 | 1 |
| wrsoil | Webb et al Soil Profile Thickness (cm) | 1.0 | 1 |
| wrroot | Webb et al Potential Storage of Water in Root Zone (mm) | 1.0 | 1 |
| wrprof | Webb et al Potential Storage of Water in Soil Profile (mm) | 1.0 | 1 |
| wrmodii | Webb et al Model II Soil Water (mm) | 1.0 | 1 |
| wrcont | Webb et al Continent Codes from the FAO/UNESCO Soil Map of the World | 1.0 | 1 |
| whcov1 | Wilson & Henderson-Sellers Primary Land Cover Classes | 1.0 | 1 |
| srztext | Staub and Rosenzweig Zobler Near-Surface Soil Texture | 1.0 | 1 |
| owe14dr | Resolution codes for OWE1.4D | 0.5 | 1 |
| owe14d | Olson World Ecosystem Classes Version 1.4D | 0.5 | 1 |
| owe13a | Olson World Ecosystems Version 1.3A | 0.5 | 1 |
| mgvc488 | 1988 MGV PCA Component 1-4 | 0.5 | 4 |
| mgv0001-12 | Average January - December Generalized Global Vegetation Index | 0.5 | 12 |
| mfwwet | Matthews and Fung Wetland Type | 1.0 | 1 |
| mfwveg | Matthews and Fung Vegetation Type | 1.0 | 1 |
| mfwsrc | Matthews and Fung Wetland Data Source | 1.0 | 1 |
| mfwsol | FAO Soil Types of Matthews & Fung Wetland Locations | 1.0 | 1 |
| mfwfrin | Matthews and Fung Fractional Inundation | 1.0 | 1 |
| maveg | Matthews Vegetation Types | 1.0 | 1 |
| malbwn | Matthews Winter Albedo (% X 100) | 1.0 | 1 |
| malbsp | Matthews Spring Albedo (% X 100) | 1.0 | 1 |
| malbsm | Matthews Summer Albedo (% X 100) | 1.0 | 1 |
| malbfa | Matthews Fall Albedo (% X 100) | 1.0 | 1 |
| macult | Matthews Cultivation Intensity | 1.0 | 1 |
| Lwtsd01-12 | Legates & Willmott January-December Temperature (std. dev.) | 0.5 | 12 |
| lwtsd00 | Legates & Willmott Annual Temperature (std. dev.) | 0.5 | 1 |
| Lwtmp01-12 | Legates & Willmott January-December Temperature (0.1C) | 0.5 | 12 |
| lwtmp00 | Legates & Willmott Annual Temperature (0.1C) | 0.5 | 1 |
| Lwmsd01-12 | Legates & Willmott January-December Measured Precipitation (std. dev.) | 0.5 | 12 |
| lwmsd00 | Legates & Willmott Annual Measured Precipitation (std. dev.) | 0.5 | 1 |
| Lwmpr01-12 | Legates & Willmott January-December Measured Precipitation (mm/month) | 0.5 | 12 |
| lwmpr00 | Legates & Willmott Annual Measured Precipitation (mm/year) | 0.5 | 1 |
| Lwcsd01-12 | Legates & Willmott January-December Corrected Precipitation (std. dev.) | 0.5 | 12 |
| lwcsd00 | Legates & Willmott Annual Corrected Precipitation (std. dev.) | 0.5 | 1 |
| Lwcpr01-12 | Legates & Willmott January-December Corrected Precipitation (mm/month) | 0.5 | 12 |
| lwcpr00 | Legates & Willmott Annual Corrected Precipitation (mm/year) | 0.5 | 1 |
| lmfmeth | Lerner et al Annual Methane Emission (Kg/Km^2) | 1.0 | 1 |
| lholdag | Leemans' Holdridge Life Zones Aggregated Classification | 0.5 | 1 |
| lhold | Leemans' Holdridge Life Zones Classification | 0.5 | 1 |
| Lctmp01-12 | Leemans and Cramer January-December Temperature (0.1C) | 0.5 | 12 |
| Lcprc01-12 | Leemans and Cramer January-December Precipitation (mm/month) | 0.5 | 12 |
| Lccld01-12 | Leemans and Cramer January-December Cloudiness (% Sunshine) | 0.5 | 12 |
| fnocwat | Navy Terrain Data--Percent Water Cover | 0.1667 | 1 |
| fnocurb | Navy Terrain Data--Percent Urban Cover | 0.1667 | 1 |
| fnocst | Navy Terrain Data--Secondary Surface Type Codes | 0.1667 | 1 |
| fnocrdg | Navy Terrain Data--Number of Significant Ridges | 0.1667 | 1 |
| fnocpt | Navy Terrain Data--Primary Surface Type Codes | 0.1667 | 1 |
| fnocmod | Navy Terrain Data--Modal Elevation (meters) | 0.1667 | 1 |
| fnocmin | Navy Terrain Data--Minimum Elevation (meters) | 0.1667 | 1 |
| fnocmax | Navy Terrain Data--Maximum Elevation (meters) | 0.1667 | 1 |
| fnocazm | Navy Terrain Data--Direction of Ridges (degrees X 10) | 0.1667 | 1 |
| etopo2 | etopo elevation | 0.0333 | 1 |

Table 2.  The variables selected for models of each of six species at a resolution of 0.05 degrees.

| Name | Description | Native Resolution | Resolution 0.05 |
|------|-------------|-------------------|-----------------|
| treecover | Continuous field data - treecover | 0.0083 | 3 |
| mgvc188 | 1988 MGV PCA Component 1 | 0.167 | 2 |
| lwmpr04 | Legates & Willmott April Measured Precipitation (mm/month) | 0.5 | 2 |
| lctmp12 | Leemans and Cramer December Temperature (0.1C) | 0.5 | 2 |
| lwmsd08 | Legates & Willmott August Measured Precipitation (std. dev.) | 0.5 | 1 |
| owe13a | Olson World Ecosystems Version 1.3A | 0.5 | 1 |
| mgvc388 | 1988 MGV PCA Component 3 | 0.5 | 1 |
| lwmsd09 | Legates & Willmott September Measured Precipitation (std. dev.) | 0.5 | 1 |
| lwmpr00 | Legates & Willmott Annual Measured Precipitation (mm/year) | 0.5 | 1 |
| lwmpr10 | Legates & Willmott October Measured Precipitation (mm/month) | 0.5 | 1 |
| lwtsd08 | Legates & Willmott August Temperature (std. dev.) | 0.5 | 1 |
| lwmpr06 | Legates & Willmott June Measured Precipitation (mm/month) | 0.5 | 1 |
| lcprc05 | Leemans and Cramer May Precipitation (mm/month) | 0.5 | 1 |
| fnocrdg | Navy Terrain Data--Number of Significant Ridges | 0.167 | 1 |
| macult | Matthews Cultivation Intensity | 1 | 1 |
| lccld06 | Leemans and Cramer June Cloudiness (% Sunshine) | 0.5 | 1 |
| lccld08 | Leemans and Cramer August Cloudiness (% Sunshine) | 0.5 | 1 |
| lwcpr02 | Legates & Willmott February Corrected Precipitation (mm/month) | 0.5 | 1 |
| mgv0001 | Average January Generalized Global Vegetation Index | 0.167 | 1 |
| lwcpr05 | Legates & Willmott May Corrected Precipitation (mm/month) | 0.5 | 1 |

Table 3. The accuracy and variables selected in the models of the Cerulean Warbler at a range of spatial resolutions.

| Resolution | In | Ex | Var 1 | Var 2 |
|---|---|---|---|---|
| 1 | 0.868 | 0.868 | lcprc04 | lmfdcow |
| 0.5 | 0.861 | 0.855 | lwcpr04 | mgvc288 |
| 0.2 | 0.878 | 0.875 | mgv0009 | lccld01 |
| 0.1 | 0.895 | 0.884 | mgv0009 | lmfpig |
| 0.1 | 0.89 | 0.886 | lcprc04 | lwtmp02 |
| 0.05 | 0.91 | 0.898 | mgv0009 | lwtmp04 |

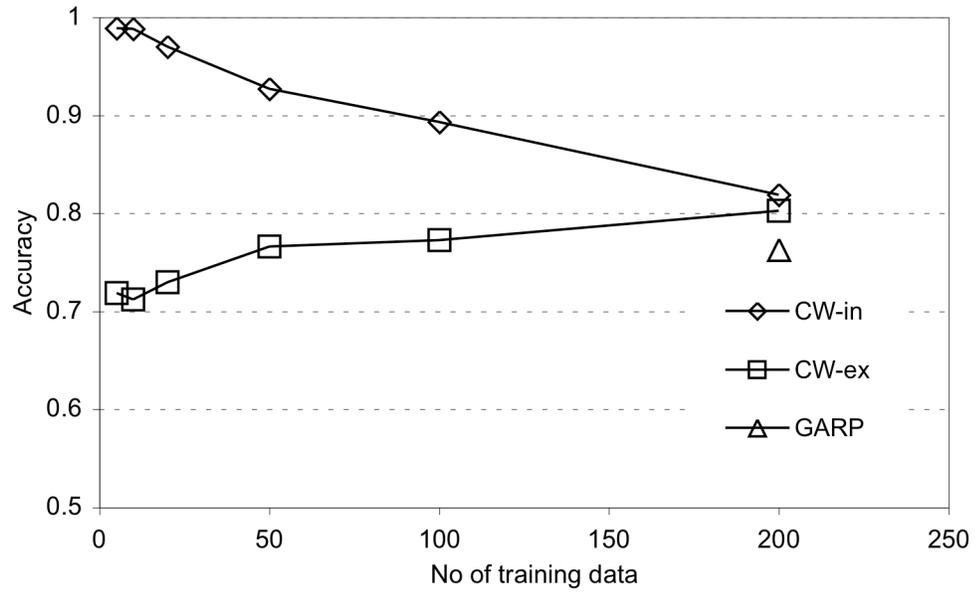

Figure 1.  Accuracy on Cerulean Warbler data with number of species data.  WW-
WhyWhere internal and external, GARP: GARP external accuracy.

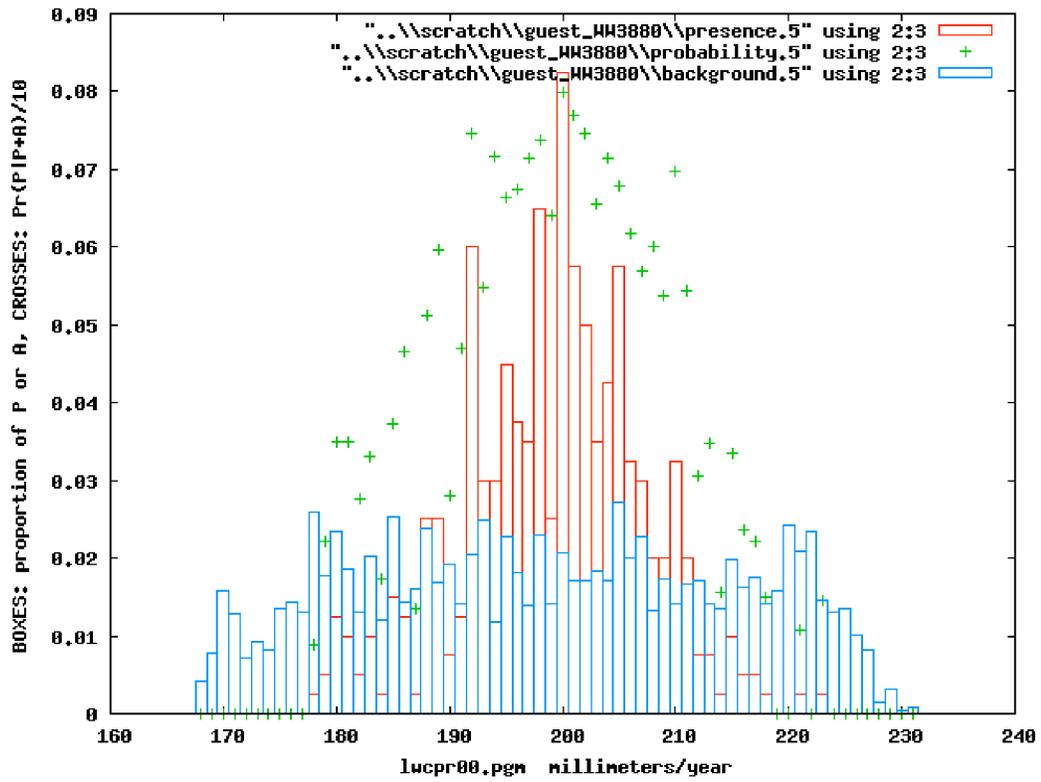

Figure 2.  The response of Cerulean Warbler to the most predictive variable.  Red line
boxes are the proportion of occurrence points and blue line boxes are the
proportion background points at each color intensity (class).  The crosses are the
probability of presence given a color class, showing a typical univariate response
to the environmental range.

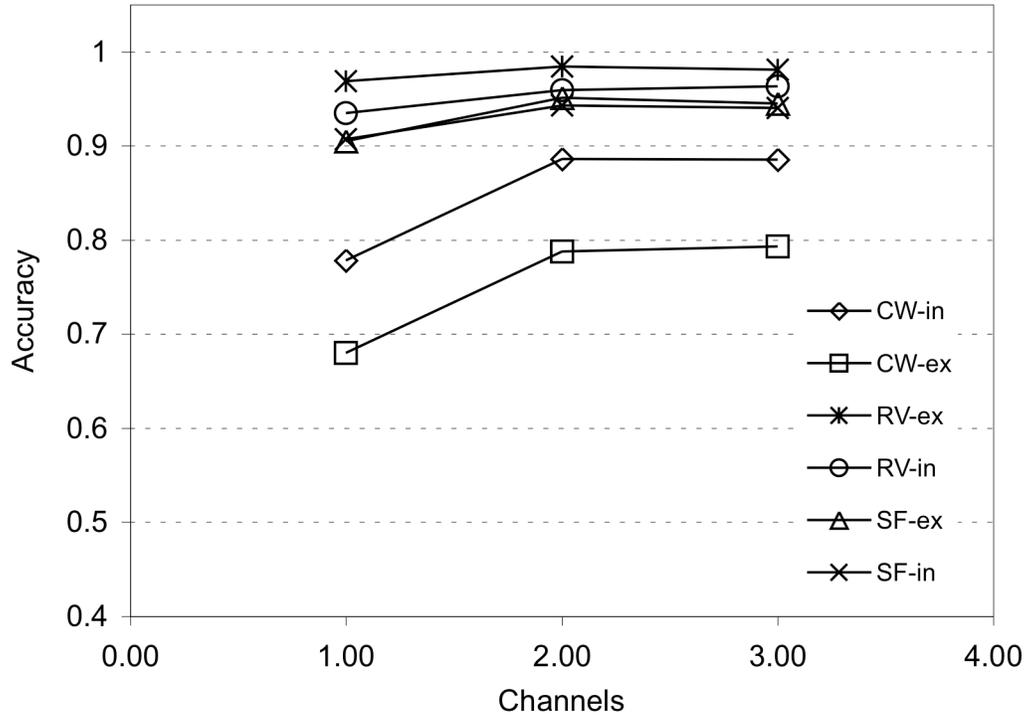

Figure 3. Increase in internal and external accuracy with number of layers in surrogate model (or iterations of algorithm) for three species and continents, CW - Cerulean Warbler (N. America), RV – Rubber Vine (Australia), and SF – Swainsons Flycatcher (S. America)

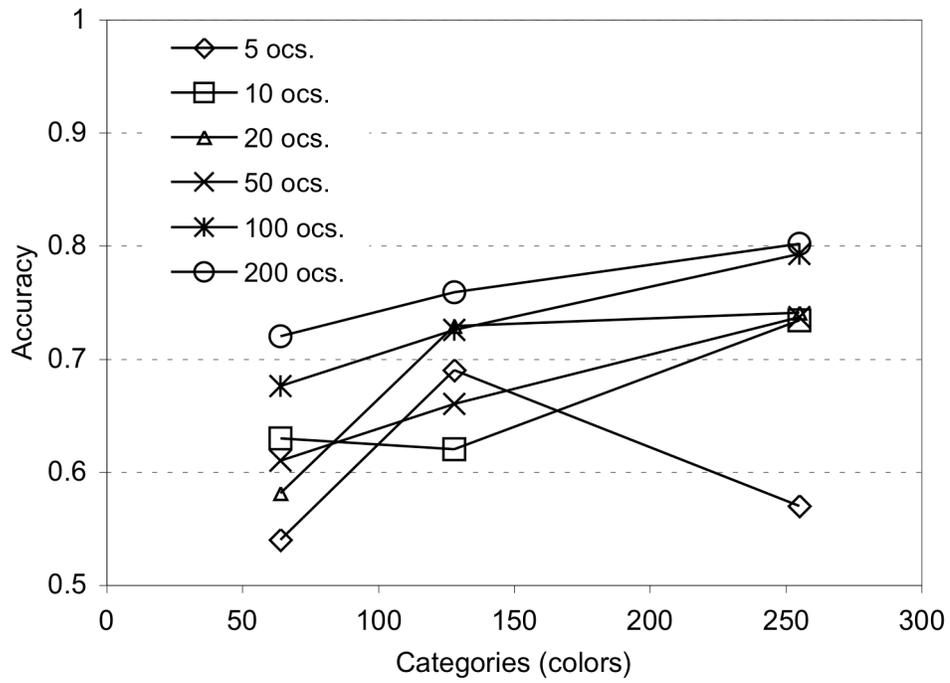

Figure 4. Variation in external accuracy of Cerulean Warbler by number of categories and training set sizes.

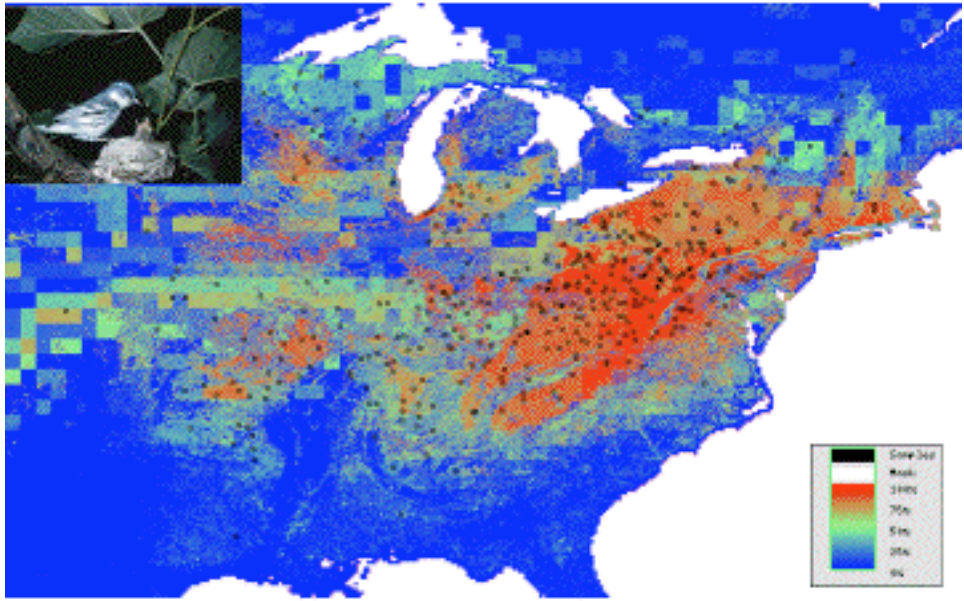

Figure 5. WhyWhere generated image of the predicted distribution of the Cerulean
        Warbler.